\documentclass[runningheads]{llncs}
\usepackage{graphicx} 

\usepackage{booktabs}
\usepackage{framed}
\usepackage{dirtytalk}
\usepackage{soul}
\usepackage{caption}

\usepackage[utf8]{inputenc} 
\usepackage[T1]{fontenc}    
\usepackage{hyperref}       
\usepackage{url}            
\usepackage{booktabs}       
\usepackage{amsfonts}       
\usepackage{amsmath}       
\usepackage{nicefrac}       
\usepackage{microtype}      
\usepackage{xcolor}         
\usepackage{graphicx}

\title{ExaRanker-Open:\\ Synthetic Explanation for IR using Open-Source LLMs}
\author{%
   Fernando Ferraretto\inst{1} \and
   Thiago Laitz\inst{1} \and
   Roberto Lotufo\inst{1,2} \and\\
   Rodrigo Nogueira\inst{1}
}

\institute{FEEC, UNICAMP, Brazil \and
NeuralMind, Brazil}

\begin{document}

\maketitle

\begin{abstract}

ExaRanker recently introduced an approach to training information retrieval (IR) models, incorporating natural language explanations as additional labels. The method addresses the challenge of limited labeled examples, leading to improvements in the effectiveness of IR models. However, the initial results were based on proprietary language models such as GPT-3.5, which posed constraints on dataset size due to its cost and data privacy. In this paper, we introduce ExaRanker-Open, where we adapt and explore the use of open-source language models to generate explanations. The method has been tested using different LLMs and datasets sizes to better comprehend the effective contribution of data augmentation. Our findings reveal that incorporating explanations consistently enhances neural rankers, with benefits escalating as the LLM size increases. Notably, the data augmentation method proves advantageous even with large datasets, as evidenced by ExaRanker surpassing the target baseline by 0.6 nDCG@10 points in our study. To encourage further advancements by the research community, we have open-sourced both the code and datasets at \url{https://github.com/unicamp-dl/ExaRanker}.
\end{abstract}

\section{Introduction and Background} \label{sec:introduction}

Previous research by Ferraretto et al.~\cite{ferraretto2023exaranker} introduced a method for training retrieval models, incorporating natural language explanations as additional labels. This approach mitigates the need for large amounts of training examples.

The method is built on the observation that a categorical label (e.g., true/false) provides limited information about the task to be learned by the AI model. This limitation makes it more challenging for the model to comprehend the subtleties of the task. It has been demonstrated that the learning process becomes more efficient when natural language explanations are provided. These explanations elucidate, for example, why a passage is relevant or not to a given query.

The ExaRanker model used GPT-3.5~\cite{ouyang2022training} in a few-shot manner to autonomaticaly augment training examples with explanations. The implementation of this method has demonstrated a reduction in the dataset size dependency, performing on par with a categorical dataset three times larger. However, the previous results suggest that the effectiveness of incorporating explanations decreases as the number of training examples increases. Additionally, the study faced limitations in the number of examples that were augmented, as it relied on a LLM exposed only via a paid API to generate explanations.

This work extends upon the ExaRanker method by leveraging open-source language models (LLMs) for generating explanations. This extension enables the method to be efficiently utilized for automatically augmenting training examples with explanations at a low cost and guaranteed data privacy.
Furthermore, we present new results using the original method on a larger dataset, confirming that explanations consistently yield benefits regardless of dataset size. We refer to the model from Ferraretto et al.~\cite{ferraretto2023exaranker} as ExaRanker-v1, and the model introduced in this paper as ExaRanker-Open.

\section{Methodology}
\label{sec:method}

In this section, the ExaRanker-Open method is presented and detailed on how it differs from the original ExaRanker.

We use the same samples randomly chosen in the prior study, consisting of 15k pairs of (query, relevant passage), and an additional 15k pairs of (query, non-relevant passage) from the training set of the MS MARCO passage ranking dataset~\cite{MSMARCOv3}. Two open-source LLMs have been used for generating explanations and augmenting the dataset: llama-2-7B-chat-hf and llama-2-70b-chat-hf~\cite{touvron2023llama}.

In addition to the dataset with 30k samples, we created a larger version containing 100k samples. This dataset is composed of 50k pairs of query-passage relevant and another 50k non-relevant pairs. In total, four versions of the dataset were generated for this experiment, as detailed in the Table~\ref{tab:Llama-ds}.

After the dataset generation, a T5-base~\cite{raffel2020exploring} model has been fine-tuned for up to 30 epochs on all dataset variations. The fine-tuning process employed the AdamW~\cite{loshchilov2018decoupled} optimizer with a learning rate set of $3e-5$, weight decay of $0.01$, and a batch size of $128$ examples ($64$ positives and $64$ negatives). Both the maximum number of input and output tokens were restricted to $512$, with sequences surpassing these limits truncated during training and inference. For comparative analysis, another T5-base model was also fine-tuned using identical hyperparameters and dataset specifications, excluding the explanations in the target text, i.e. using categorial labels only. The finetuning method for this baseline model is similar to one used by monoT5~\cite{nogueira2020document}, except that here we use smaller versions of the MS MARCO dataset and a different optimizer.

\begin{table}
\centering
\begin{tabular}{p{1.0in}p{0.6in}p{0.6in}}
\toprule
\textbf{LLM model} & \textbf{Relevant Samples} & \textbf{Total Samples} \\
\midrule
\texttt Llama-2-7B & 15k & 30k \\
\texttt Llama-2-7B & 50k & 100k \\
\midrule
\texttt Llama-2-70B & 15k & 30k \\
\texttt Llama-2-70B & 50k & 100k \\

\bottomrule
\end{tabular}
\vspace{0.1cm}
\caption{Datasets generated with the open-source LLMs.}
\label{tab:Llama-ds}
\end{table}

Besides the ExaRanker-Open method as described before, the prior ExaRanker-v1 experiment was expanded to encompass a 20 times larger dataset of 300k relevant pairs of query-passage and 300k non-relevant pairs. Following the same methodology, samples were randomly selected, and utilizing the few-shot prompt, explanations were generated by GPT-3.5-turbo. We used greedy decoding and restricting the output to 256 tokens.

The next step is to  fine-tuned a T5-base model over 30 epochs with explanations. The experiment was conducted across eight different dataset sizes: 2.5k, 5k, 10k, 15k, 50k, 100k, 150k, and 300k relevant pairs of query-passage. In all cases, an equal number of non-relevant passages were employed, resulting in the smallest dataset comprising 5k samples and the largest reaching 600k. As done before, another monoT5 was fine-tuned within the same dataset but without explanations. This model serves as a benchmark for comparison, enabling an evaluation of the benefits derived from using explanations to augment the dataset and fine-tune the neural ranker in eight different dataset sizes.

To evaluate the advantages of incorporating explanations, each fine-tuned model was evaluated in a zero-shot manner across six datasets from the BEIR benchmark~\cite{thakur2021beir}: Robust04~\cite{robust04}, TREC-COVID~\cite{trec-covid}, DBPedia~\cite{hasibi2017dbpedia}, FiQA~\cite{fiqa}, TREC-NEWS~\cite{soboroff2018trec}, and NFCorpus~\cite{boteva2016full}. TREC-DL 2020~\cite{dl20} serves as a validation set for selecting the best checkpoint. Different attempts were made for each experiment, with this study reporting the average results obtained.

\section{Results}

The results obtained with open-source LLMs are shown in the Table~\ref{tab:main_os}. The initial row represents the T5-base model fine-tuned without data augmentation, solely relying on categorical labels. The second row reflects the earlier results of ExaRanker-v1 using GPT-3.5-turbo. The last two rows present the outcomes of this study, employing Llama-2-70B and Llama-2-7B, respectively.

\begin{figure}
    \centering
    \includegraphics[scale=0.60]{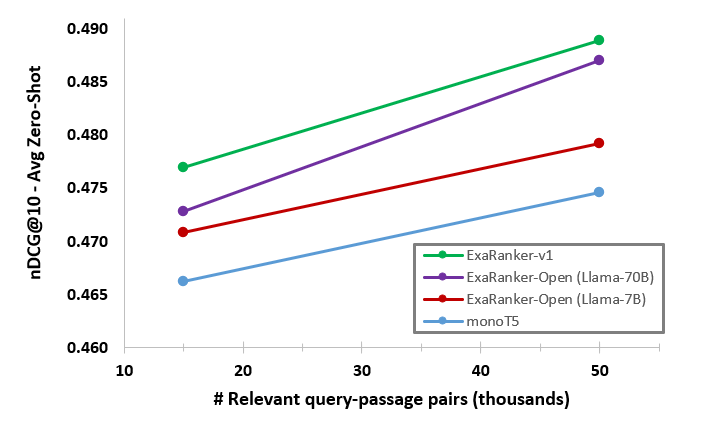}
    \caption{Average zero-shot results on 6 datasets of the BEIR benchmark with respect to training dataset size, comparing the 4 models evaluated.}
    \label{fig:results-OS}
\end{figure}

As evident, the zero-shot effectiveness is enhanced when employing a larger LLM model. Illustrated in Figure~\ref{fig:results-OS}, the performance of Llama-2-7B surpasses that of the monoT5 without explanations in both evaluated dataset sizes, 15k and 50k relevant pairs of query-passage. However, Llama-2-70B outperforms Llama-2-7B, specially when using the larger dataset with 50k query-relevant passage pairs. Ultimately, ExaRanker-v1, using GPT-3.5, remains the top-performing model.

These results strongly suggest the quality of data augmentation produced by each Llama model size. As expected, larger models exhibit superior natural language processing capabilities, leading to a more substantial extraction of signals that can be effectively utilized during the fine-tuning phase.

\begin{table*}[] \small
\caption{Results (nDCG@10) of open-source LLMs. Average zero-shot (all except DL 20). The column ``Ft Pos.'' is the number of positive training examples on which the model was finetuned.}
\vspace{0.1cm}
\centering\centering\resizebox{1.0\textwidth}{!}{
\begin{tabular}{lcc|ccccccccc}
\toprule
\textbf{Model} & \textbf{LLM} & \textbf{Ft Pos.} & \textbf{DL 20} & \textbf{Robust} & \textbf{Covid} & \textbf{Dbp} & \textbf{FiQA} & \textbf{News} & \textbf{NFC} &
\textbf{Avg ZS}
\\
\midrule

monoT5 & n/a & 15k & 0.656 & 0.523 & 0.746 & 0.392 & 0.382 & 0.409 & 0.344 & 0.466

\\
\texttt{ExaRanker-v1} & GPT-3.5 &  15k & 0.683 & 0.531 & 0.752 & 0.403 & 0.408 & 0.415 & 0.352 & 0.477
\\
\texttt{ExaRanker-Open} & Llama-2-70B &  15k & 0.653 & 0.551 & 0.730 & 0.398 & 0.393 & 0.425 & 0.341 & 0.473
\\
\texttt{ExaRanker-Open} & Llama-2-7B & 15k & 0.662 & 0.523 & 0.737 & 0.407 & 0.392 & 0.421 & 0.344 & 0.471

\\
\midrule

monoT5 & n/a & 50k & 0.653 & 0.534 & 0.757 & 0.384 & 0.396 & 0.426 & 0.350 & 0.475

\\
\texttt{ExaRanker-v1} & GPT-3.5 & 50k & 0.673 & 0.540 & 0.778 & 0.423 & 0.413 & 0.431 & 0.349 & 0.489

\\
\texttt{ExaRanker-Open} & Llama-2-70B & 50k & 0.670 & 0.563 & 0.757 & 0.414 & 0.403 & 0.440 & 0.345 & 0.487

\\
\texttt{ExaRanker-Open} & Llama-2-7B & 50k & 0.670 & 0.529 & 0.741 & 0.419 & 0.398 & 0.439 & 0.349 & 0.479

\\

\bottomrule
\end{tabular}
}
\vspace{0.1cm}

\vspace{-0.2cm}

\label{tab:main_os}
\end{table*}

The results of the second experiment, an extension of ExaRanker-v1, are shown in the Table~\ref{tab:main_exav1}. Contrary to the results presented in the original ExaRanker paper, here we observe that data augmentation consistently yields benefits, even with large datasets. On average, the effectiveness of ExaRanker-v1 is 1.4 nDCG@10 points higher than that of the monoT5 trained without explanations.

\begin{figure}
    \centering
    \includegraphics[scale=0.50]{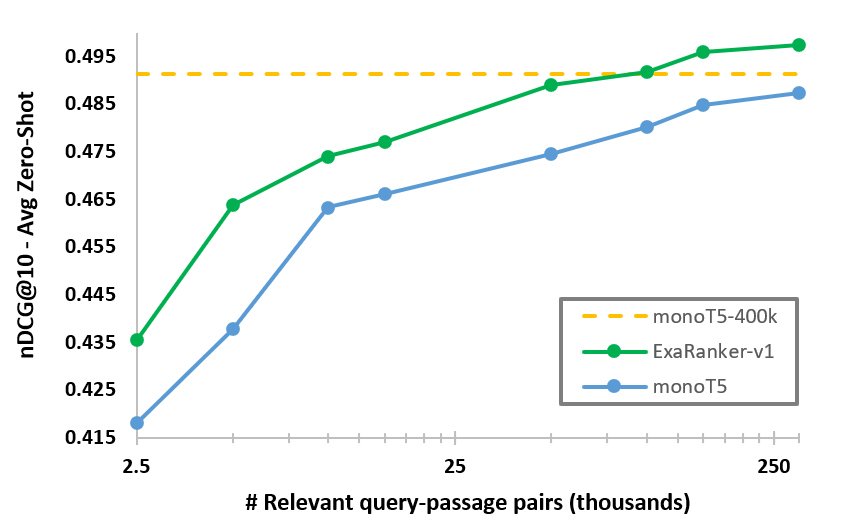}
    \caption{Average zero-shot results on 6 datasets of the BEIR benchmark. monoT5-400k is finetuned on the 400k relevant query-passage pairs from MS MARCO without explanations. Note the log scale in horizontal axis.}
    \label{fig:results-large}
\end{figure}

Indeed, the model introduced by our method outperforms the strong monoT5-400k baseline, by 0.8 NDCG@10 points, as illustrated in Figure~\ref{fig:results-large}. This baseline is the monoT5-base model, which was trained over the entire MS MARCO dataset, consisting of 400k relevant pairs. In contrast, our model is trained on a dataset with 300k relevant pairs but still achieves superior performance.

\begin{table*}[] \small
\caption{Results (nDCG@10) of larger datasets. Average zero-shot (all except DL 20). The column ``Ft Pos.'' is the number of positive training examples on which the model was finetuned. The "*" marker denotes the results from Ferraretto et al.~\cite{ferraretto2023exaranker}.}
\vspace{0.1cm}
\centering
\begin{tabular}{cccccccccccc}
\toprule
\textbf{Model} & \textbf{Ft Pos.} & \textbf{DL 20} & \textbf{Robust} & \textbf{Covid} & \textbf{Dbp} & \textbf{FiQA} & \textbf{News} & \textbf{NFC} &
\textbf{Avg ZS}
\\
\midrule
\texttt{monoT5} & 300k & 0.662 & 0.532 & 0.780 & 0.412 & 0.403 & 0.446 & 0.350 & 0.487

\\
\texttt{ExaRanker-v1} & 300k & 0.682 & 0.558 & 0.784 & 0.427 & 0.416 & 0.451 & 0.349 & 0.497

\\

\midrule
\texttt{monoT5} & 150k & 0.663 & 0.537 & 0.790 & 0.395 & 0.396 & 0.443 & 0.349 & 0.485

\\
\texttt{ExaRanker-v1} & 150k & 0.684 & 0.559 & 0.781 & 0.426 & 0.413 & 0.447 & 0.348 & 0.496

\\

\midrule
\texttt{monoT5} & 100k & 0.658 & 0.528 & 0.774 & 0.400 & 0.396 & 0.434 & 0.350 & 0.480

\\
\texttt{ExaRanker-v1} & 100k & 0.677 & 0.552 & 0.776 & 0.419 & 0.412 & 0.440 & 0.350 & 0.492

\\

\midrule
\texttt{monoT5} & 50k & 0.653 & 0.534 & 0.757 & 0.384 & 0.396 & 0.426 & 0.350 & 0.475

\\
\texttt{ExaRanker-v1} & 50k & 0.673 & 0.540 & 0.778 & 0.423 & 0.413 & 0.431 & 0.349 & 0.489

\\

\midrule
\texttt{monoT5} & 15k & 0.656 & 0.523 & 0.746 & 0.392 & 0.382 & 0.409 & 0.344 & 0.466

\\
\texttt{ExaRanker-v1} & 15k & 0.683 & 0.531 & 0.752 & 0.403 & 0.408 & 0.415 & 0.352 & 0.477

\\

\midrule
\texttt{monoT5$^*$} & 10k & 0.643 & 0.510 & 0.749 & 0.379 & 0.374 & 0.426 & 0.341 & 0.463
\\
\texttt{ExaRanker-v1$^*$} & 10k & 0.667 & 0.527 & 0.752 & 0.409 & 0.393 & 0.418 & 0.347 & 0.474
\\

\midrule
\texttt{monoT5$^*$} & 5k & 0.625 & 0.488 & 0.693 & 0.364 & 0.337 & 0.417 & 0.328 & 0.438
\\
\texttt{ExaRanker-v1$^*$} & 5k & 0.665 & 0.505 & 0.750 & 0.389 & 0.380 & 0.414 & 0.345 & 0.464
\\

\midrule
\texttt{monoT5$^*$} & 2.5k & 0.611 & 0.486 & 0.666 & 0.334 & 0.328 & 0.370 & 0.325 & 0.418
\\
\texttt{ExaRanker-v1$^*$} & 2.5k & 0.650 & 0.496 & 0.686 & 0.393 & 0.306 & 0.398 & 0.335 & 0.436
\\

\bottomrule
\end{tabular}
\vspace{0.1cm}

\vspace{-0.2cm}

\label{tab:main_exav1}
\end{table*}

\section{Conclusion}

This study introduces ExaRanker-Open, an enhanced iteration of ExaRanker-v1~\cite{ferraretto2023exaranker} that uses open-source LLMs to augment categorical datasets with natural language explanations. The findings demonstrate the advantages of this approach within IR domain, revealing enhancement in the effectiveness of the models fine-tuned through this methodology. Furthermore, experiments on larger training datasets validates the advantages of data augmentation when employing a neural ranker for IR tasks. ExaRanker outperforms a strong baseline trained on a larger dataset but without natural language explanations associated in the target.


\section*{Acknowledgments}

This research was partially funded by grant 2022/01640-2 from Fundação de Amparo à Pesquisa do Estado de São Paulo (FAPESP). Roberto Lotufo is partially supported by CNPq (The Brazilian National Council for Scientific and Technological Development) under grant 313047/2022-7.

\bibliographystyle{abbrv}

\bibliography{refs}

\end{document}